\documentclass[
preprint,
superscriptaddress,prd,tightenlines,
nofootinbib,eqsecnum,
amsfonts,amsmath,amssymb]{revtex4}

\usepackage{bm}
\usepackage{graphics,graphicx,epsfig,color,ulem}
\usepackage{subfigure}

\newcommand{\ud}{\mathrm{d}}

\begin{document}

\title{Does an atom interferometer test the gravitational redshift \\
  at the Compton frequency\,?}

\author{Peter Wolf} \affiliation{LNE-SYRTE, CNRS UMR 8630, UPMC,
  Observatoire de Paris, 61 avenue de l'Observatoire, 75014 Paris,
  France} \author{Luc Blanchet} \affiliation{GRECO, Institut
  d'Astrophysique de Paris, CNRS UMR 7095, UPMC, 98$^\text{bis}$
  boulevard Arago, 75014 Paris, France} \author{Christian J. Bord\'e}
\affiliation{LNE-SYRTE, CNRS UMR 8630, UPMC, Observatoire de Paris, 61
  avenue de l'Observatoire, 75014 Paris, France}
\affiliation{Laboratoire de Physique des Lasers, CNRS UMR 7538, 99
  avenue Jean-Baptiste Cl\'ement, 93430 Villetaneuse, France}
\author{Serge Reynaud} \affiliation{Laboratoire Kastler Brossel, CNRS
  UMR 8552, ENS, UPMC, Campus Jussieu, 75252 Paris, France}
\author{Christophe Salomon} \affiliation{Laboratoire Kastler Brossel
  et Coll\`ege de France, CNRS UMR 8552, ENS, UPMC, 24 rue Lhomond,
  75231 Paris, France} \author{Claude Cohen-Tannoudji}
\affiliation{Laboratoire Kastler Brossel et Coll\`ege de France, CNRS
  UMR 8552, ENS, UPMC, 24 rue Lhomond, 75231 Paris, France}
\date{\today}

\begin{abstract}
  Atom interferometers allow the measurement of the acceleration of
  freely falling atoms with respect to an experimental platform at
  rest on Earth's surface. Such experiments have been used to test the
  universality of free fall by comparing the acceleration of the atoms
  to that of a classical freely falling object. In a recent paper,
  M\"uller, Peters and Chu [A precision measurement of the
  gravitational redshift by the interference of matter waves, Nature
  {\bf 463}, 926-929 (2010)] argued that atom interferometers also
  provide a very accurate test of the gravitational redshift (or
  universality of clock rates). Considering the atom as a clock
  operating at the Compton frequency associated with the rest mass,
  they claimed that the interferometer measures the gravitational
  redshift between the atom-clocks in the two paths of the
  interferometer at different values of gravitational potentials. In
  the present paper we analyze this claim in the frame of general
  relativity and of different alternative theories. We show that the
  difference of ``Compton phases'' between the two paths of the
  interferometer is actually zero in a large class of theories,
  including general relativity, all metric theories of gravity, most
  non-metric theories and most theoretical frameworks used to
  interpret the violations of the equivalence principle. Therefore, in
  most plausible theoretical frameworks, there is no redshift effect
  and atom interferometers only test the universality of free fall. We
  also show that frameworks in which atom interferometers would test
  the redshift pose serious problems, such as (i) violation of the
  Schiff conjecture, (ii) violation of the Feynman path integral
  formulation of quantum mechanics and of the principle of least
  action for matter waves, (iii) violation of energy conservation, and
  more generally (iv) violation of the particle-wave duality in
  quantum mechanics. Standard quantum mechanics is no longer valid in
  such frameworks, so that a consistent interpretation of the
  experiment would require an alternative formulation of quantum
  mechanics. As such an alternative has not been proposed to date, we
  conclude that the interpretation of atom interferometers as testing
  the gravitational redshift at the Compton frequency is unsound.
\end{abstract}

\maketitle

\section{Introduction}\label{I}

\subsection{Motivation}\label{IA}

General relativity (GR) has historically been underpinned by distinct
types of experimental measurements, which have tested the geometric
nature of the theory as well as the specific field equations of GR. In
particular, the four ``classical tests'' are still carried out today
with continuously improving accuracy. The first one is the observation
of the perihelion advance of planets, for example Mercury, which had
been known by Le Verrier in 1845, and explained by general relativity
in 1915 (see \cite{Chazy} for a review). The second is the bending of
light by the Sun, which was predicted by general relativity and first
measured by Eddington in 1919 during a solar eclipse
\cite{Bertotti}. Another test is the relativistic time delay of radio
waves grazing the Sun during their round trip to Mercury and which was
computed and measured by Shapiro in 1964 \cite{Shapiro}.

The gravitational redshift, also considered as a ``classical test'',
is actually a test of one facet of the equivalence principle. The weak
equivalence principle has been verified with high precision using
torsion balances \cite{Dicke,Braginsky,Schlamminger} and the Lunar
laser ranging \cite{Williams}. The gravitational redshift was
predicted by Einstein but not observed for a long time
\cite{Bertotti}. It became observable with the advent of high
precision quantum spectroscopy and was first measured in 1960 by Pound
and Rebka \cite{Pound} (see also \cite{PoundSnider}), who used gamma
ray spectroscopy of the radiation emitted and absorbed by ${}^{57}$Fe
nuclei. The emitter and absorber were placed at the top and bottom of
a $22.5\,\mathrm{m}$ high tower at Harvard and the frequency
difference predicted by GR was measured with about $1\%$
uncertainty. Since then, the initial prediction of a gravitational
redshift of solar spectral lines emitted at the Sun's surface has also
been measured \cite{Lopresto}.

In a test of the gravitational redshift using clocks, one checks
that the clock rates are universal --- i.e. the relative rates
depend only on the difference of gravitational potentials (as
determined by the trajectories of massive test bodies) but not on
the nature and internal structure of the clocks. In 1976 a
hydrogen-maser clock was launched on a rocket to an altitude of
$10,000\,\mathrm{km}$ and its frequency compared to a similar
clock on ground. This yielded a test of the gravitational redshift
with about $10^{-4}$ accuracy \cite{Vessot}. The European Space
Agency will fly in 2013 on the International Space Station the
Atomic Clock Ensemble in Space (ACES), including a highly stable
laser-cooled atomic clock, which will test (in addition to many
other applications in fundamental physics and metrology) the
gravitational redshift to a precision of about $10^{-6}$
\cite{Cacciapuoti}. The gravitational redshift is also tested in
null redshift experiments in which the rates of different clocks
(based on different physical processes or different atoms) are
compared to each other \cite{Turneaure,Blatt}.

Quite generally in a modern context, tests of GR measure the
difference between the predictions of GR and of some generalized
alternative theory or theoretical framework (see \cite{Will} for a
review). For example, GR may be compared to alternative metric
theories of gravitation such as the Jordan-Brans-Dicke scalar-tensor
theory \cite{Jordan,Brans}. The metric theories may also be compared
with non-metric theories of gravity such as the Belinfante-Swihart
theory \cite{BelinfanteS}. It has to be kept in mind that the
classification and inter-comparison of different tests have then to be
defined with respect to the framework used.

In the following we will investigate, whether a gravimeter (or
accelerometer) based on atom interferometry, can test the
gravitational redshift. Atom interferometers allow one to measure the
acceleration of atoms falling in the gravitational field of the Earth
(with respect to the experimental platform at rest on Earth). The
beam-splitter for atom waves is realized through the interaction of
atoms with laser beams resonant with a hyperfine atomic transition. It
is known
\cite{Borde89,Kasevich,Storey,Wolf,Borde01,Borde02,Borde04,Borde08}
that the main contribution to the phase shift in atom interferometers
comes from the phase imprinted on the matter wave by the beam
splitters. Following earlier suggestions by Bord\'e \cite{Borde89},
the first gravimeter based on atom interferometry was realized by
Kasevich and Chu \cite{Kasevich}. Before that, the first gravimeter
based on neutron interferometry had been realized by Colella,
Overhauser and Werner \cite{COW} (see \cite{Greenberger} for a
review).

Atom interferometers have reached high sensitivities in the
measurement of the gravitational acceleration
\cite{Peters,refsyrte} and rotational acceleration through the
Sagnac effect \cite{Riehle,refsagnac1,refsagnac2}. This yields
very important tests of the weak equivalence principle or
universality of free fall (UFF) when comparing the free fall of
atoms with that of classical macroscopic matter (in practice a
nearby freely falling corner cube whose trajectory is monitored by
lasers). The relative precision of such tests of the UFF is
currently $7\times10^{-9}$, using Cs \cite{Peters,Mueller} or Rb
\cite{refsyrte} atoms. Although it remains less sensitive than
tests using macroscopic bodies of different composition which have
reached a precision of $2\times10^{-13}$
\cite{Williams,Schlamminger}, this UFF test is interesting as it
is the most sensitive one comparing the free fall of quantum
objects (namely C{\ae}sium atoms) with that of a classical test
mass (the corner cube).

\subsection{Overview}\label{IB}

In a recent paper, M\"uller, Peters and Chu \cite{Mueller} (hereafter
abbreviated as MPC) proposed a new interpretation of atom
interferometry experiments as testing the gravitational redshift, that
is also the universality of clock rates (UCR), with a precision
$7\times10^{-9}$, which is several orders of magnitude better than the
best present \cite{Vessot} and near future \cite{Cacciapuoti} clock
tests.

The main argument of MPC (see also the more detailed papers
\cite{Hohensee,Hohensee2}) is based on an analogy between atom
interferometry experiments and classical clock experiments. The
idea of clock experiments is to synchronize a pair of clocks when
they are located closely to one another, and move them to
different elevations in a gravitational field. The gravitational
redshift will decrease the oscillation frequency of the lower
clock relative to the higher one, yielding a measurable phase
shift between them. There are two methods for measuring the
effect. Either we bring the clocks back together and compare the
number of elapsed oscillations, or we measure the redshift by
means of continuous exchanges of electromagnetic signals between
the two clocks. In both methods one has to monitor carefully the
trajectories of the two clocks. For example, in the second method
one has to remove the Doppler shifts necessarily appearing in the
exchanges of electromagnetic signals.

In the first method, the phase difference between the two clocks when
they are recombined together, can be written as a difference of
integrals over proper time,
\begin{equation}\label{phase_clock}
  \Delta \varphi_\text{clock} = \omega \left[\int_{\text{I}} \ud\tau -
  \int_{\text{II}} \ud\tau \right] \equiv \omega \oint \ud\tau\,.
\end{equation}
The two clocks have identical proper frequency $\omega$. We denote by
I and II the two paths (with say I being at a higher altitude, i.e. a
lower gravitational potential) and use the notation $\oint\ud\tau$ to
mean the difference of proper times between the two paths, assumed to
form a close contour. The integrals in \eqref{phase_clock} are
evaluated along the paths of the clocks, and we may use the
Schwarzschild metric to obtain an explicit expression of the measured
phase shift in the gravitational field of the Earth.

It is indeed true that the phase shift measured by an atom
interferometer contains a contribution which is similar to the clock
phase shift \eqref{phase_clock}. In this analogy, the role of the
clock's proper frequency is played by the atom's (de Broglie-)Compton
frequency \cite{deBroglie}
\begin{equation}\label{compton}
\omega_\text{C} = \frac{m c^2}{\hbar}\,,
\end{equation}
where $m$ denotes the rest mass of the atom. However the phase shift
includes also another contribution $\Delta \varphi_\ell$ coming from
the interaction of the laser light used in the beam-splitting process
with the atoms. Thus,
\begin{equation}\label{phase_interf}
\Delta \varphi = \omega_\text{C} \oint \ud\tau + \Delta \varphi_\ell\,.
\end{equation}
Here we are assuming that the two paths close up at the entry and exit
of the interferometer; otherwise, additional terms have to be added to
\eqref{phase_interf}. The first term in \eqref{phase_interf} is
proportional to the atom's mass through the Compton frequency and
represents the difference of Compton phases along the two classical
paths. In contrast, the second term $\Delta \varphi_\ell$ does not
depend on the mass of the atoms.

At first sight, the first term in Eq.~\eqref{phase_interf} could
be used for a test of the gravitational redshift, in analogy with
classical clock experiments, cf. Eq.~\eqref{phase_clock}, and the
precision of the test could be very good, because the Compton
frequency of the C{\ae}sium atom is very high, $\omega_\text{C}
\approx 2 \pi \times 3.0 \times 10^{25}\,\text{Hz}$. However, as
shown in a previous brief reply \cite{WolfNat},\footnote{See also
    \cite{MuellerNat} for MPC's answer to our reply.} this
re-interpretion of the atom interferometer as testing the UCR is
fundamentally incorrect.

The clear-cut argument showing that the atom
interferometer does not measure the redshift is that the
``atom-clock'' contribution, i.e.  the first term in
\eqref{phase_interf}, is in fact \textit{zero} for a closed total path
\cite{Storey,Borde01,Borde04,Borde08}. Thus only the second term
$\Delta \varphi_\ell$ remains in Eq.~\eqref{phase_interf}. The final
phase shift,
\begin{equation}\label{phase_total}
\Delta \varphi = \Delta \varphi_\ell = k \,g \,T^2\,,
\end{equation}
depends on the wavevector $k$ of the lasers, on the interrogation time
$T$ and on the local gravity $g$, but is independent of the Compton
frequency \eqref{compton}.

In the present paper, we expand our previous brief reply and address
in detail the involved issues. As we shall detail below, the first
term in \eqref{phase_interf} vanishes in GR and in all metric theories
of gravity, and in a large class of non-metric test theories which
emcompasses most theoretical frameworks used to interpret the
violations of the different facets of the equivalence principle
\cite{Will}.

The key point about the result \eqref{phase_total} is a consistent
calculation of the two paths in the atom interferometer and of the
phases along these paths, both derived from the same classical
action, using in a standard way the principle of least action. At
the deepest level, the principle of least action and its use in
atom interferometry comes from the Feynman path integral
formulation of quantum mechanics or equivalently the Schr\"odinger
equation \cite{Storey}.

We will see in the following that the argument of MPC (developed
in the initial paper \cite{Mueller}) implies a violation of the
principle of least action. The significance and the sensitivity of
atom interferometry should then be evaluated in the framework of
an alternative theory because it is not consistent to calculate,
as MPC do in \cite{Mueller},\footnote{See ``Methods'' in
\cite{Mueller}, where they consider two scenarios. The first one
clearly states that the trajectories are not modified whilst the
atomic phases are. The second uses $g'$ for the trajectories and
$g(1+\beta)$ for the phases, which again corresponds to two
different Lagrangians.} the dephasing of the interferometer by
using the Feynman approach with two different Lagrangians.
Instead, a different approach for the description of atom
interferometers has to be developed for dealing with the violation
of fundamental principles of quantum mechanics.

These investigations lie at the interface between quantum mechanics
and gravitational theories. One could wonder whether it is possible to
provide clear answers to the question asked in the title of this
paper. We shall see below that a number of clear-cut statements and
definite conclusions will be obtained, provided the theoretical
frameworks are specified clearly and unambiguously. To this aim, we
shall consider two general alternative theoretical frameworks.

The first framework, which we call the ``modified Lagrangian
formalism'', consists of a deformation of the Lagrangian of GR
\cite{Nordtvedt,Haugan,Will}. Such a modified Lagrangian formalism
is at the basis of the usual interpretation of the various tests
of the equivalence principle, including the tests of the UCR and
the UFF, in a way consistent with Schiff's conjecture
\cite{Schiff}, the principle of least action and the basic
principles of quantum mechanics. Within this general formalism UCR
and UFF tests are related (violation of one implies violation of
the other) with the quantitative relationship depending on the
formalism used. As we will show, in such theories the atom
interferometer phase shift arising from the free evolution of the
atoms is rigorously zero and thus the Compton frequency plays no
role. Instead, the measured phase shift comes from the laser
interactions and measures the free fall of the atoms using the
laser as a ``ruler''. In that sense the atom interferometers test
UFF, and indeed appear as such in the more recent publication of
MPC \cite{Hohensee2}.\footnote{In \cite{Hohensee2} atom
interferometer tests set limits on the same parameter combination
as classical UFF tests, namely $\beta_1+\xi^{\rm
bind}_1-\beta_2-\xi^{\rm bind}_2$ (\cite{Hohensee2}, p.~4), where
the subscripts refer to test masses 1 and 2 (e.g. $1=$Cs and
$2=$falling corner cube in \cite{Peters} whilst $1=$Ti and $2=$Be
in \cite{Schlamminger}), and $\xi^{\rm bind}$ refers to the
nuclear binding energy of the test masses. On the other hand UCR
tests in \cite{Hohensee2} set limits on either $\beta_1-\xi^{\rm
trans}_2$ or $\xi^{\rm trans}_1-\xi^{\rm trans}_2$
(\cite{Hohensee2}, p.~3), where $\xi^{\rm trans}$ is related to
the atomic transition of the atom used in the clock and not to its
nuclear binding energy.}

In the second framework, which we call the ``multiple Lagrangian
formalism'', the \textit{motion} of test particles (atoms or
macroscopic bodies) obeys the standard GR Lagrangian in a
gravitational field, whereas the \textit{phase} of the
corresponding matter waves obeys a \textit{different} Lagrangian.
The MPC analysis in \cite{Mueller} belongs to this framework which
raises extremely difficult problems. It violates the Feynman path
integral method which is at the basis of the computation of the
quantum phase shift, as well as the Schiff conjecture. In
particular, the phase shift is derived in a manner which is not
consistent with standard quantum mechanics. This derivation would
thus require the development of a mathematically sound theory
allowing one to go beyond standard quantum mechanics. It is only
after this development that the interpretation of atom
interferometry experiments as testing the gravitational redshift
could be attributed an unambiguous meaning.

At this point, we also want to mention other crucial differences
between atom interferometry and clock experiments. In clock
experiments the trajectories of the clocks are continuously controlled
for instance by continuous exchange of electromagnetic signals. In
atom interferometry in contrast, the trajectories of the atoms are not
measured independently but theoretically derived from the Lagrangian
and initial conditions. Furthermore, we expect that it is impossible
to determine independently the trajectories of the wave packets
without destroying the interference pattern at the exit of the
interferometer.\footnote{Note also that the phase shift
  \eqref{phase_clock} for clocks is valid in any gravitational field,
  with any gravity gradients, since it is simply the proper time
  elapsed along the trajectories of the clocks in a gravitational
  field. By contrast, the phase shift \eqref{phase_interf} is known
  only for quadratic Lagrangians (see Sec.~\ref{IIA}) and cannot be
  applied in a gravitational field with large gravity gradients, or
  more generally with any Lagrangian that is of higher order.}

Another important difference lies in the very notion of a clock.
Atomic clocks use the extremely stable energy difference between
two internal states. By varying the frequency of an interrogation
signal (e.g. microwave or optical), one obtains a resonant signal
when the frequency is tuned to the frequency of the atomic
transition. In their re-interpretation, MPC view the entire atom
as a clock ticking at the Compton frequency associated with its
rest mass. But the ``atom-clock'' is not a real clock in the
previous sense, since it does not deliver a physical signal at
Compton frequency, as also recently emphasized in the same context
in Ref.~\cite{Samuel}.

The plan of this paper is as follows. In Sec.~\ref{II} we discuss the
prediction of GR, emphasizing its foundations in Secs.~\ref{IIA} and
\ref{IIB}, and standard interpretation in Sec.~\ref{IIC}. In
Sec.~\ref{III} we investigate the alternative multiple Lagrangian
formalism (Sec.~\ref{IIIA}) and modified Lagrangian formalism
(Sec.~\ref{IIIB}). Our conclusions are summarized in
Sec.~\ref{IV}. The case of neutron interferometry is briefly discussed
in Appendix \ref{appA}. We provide some estimate of the influence of
high-order gravity gradients in Appendix \ref{appB}.

\section{Gravimetry using atom interferometers}\label{II}

\subsection{Path integral formalism}\label{IIA}

Theoretical tools for atom optics and interferometry have been
extensively developed
\cite{Storey,Borde01,Borde02,Borde04,Borde08,Dimopoulos}. Since
atom interferometers are close to the classical regime
($S\gg\hbar$), a path integral approach is very appropriate since
it reduces to a calculation of integrals along classical paths for
Lagrangians which are at most quadratic in position and velocity.
The path integral approach to matter wave interferometry is
developed in detail in Ref.~\cite{Storey}. Equivalent results have
been obtained using non-relativistic or relativistic wave
equations of quantum mechanics (see e.g. Ref.~\cite{Borde01} for a
review). Here we only recall the points relevant to our
discussion. We denote by $K(z_b,t_b;z_a,t_a)$ the quantum
propagator, that provides the wavefunction of the atom at point
$(z_b,t_b)$ given the wavefunction at any initial point
$(z_a,t_a)$,
\begin{equation}\label{fonctiononde}
\psi(z_b,t_b) = \int \ud z_a\,K(z_b,t_b;z_a,t_a)\,\psi(z_a,t_a)\,,
\end{equation}
where the integration is over all initial positions $z_a$ at the given
time $t_a$. In Feynman's formulation the propagator is a sum over all
possible paths $\Gamma$ connecting $(z_a,t_a)$ to $(z_b,t_b)$,
\begin{equation}\label{propagateur}
K(z_b,t_b;z_a,t_a) = \int
\mathcal{D}z_\Gamma(t)\,\text{e}^{\frac{\rm{i}}{\hbar}
S_\Gamma}\,,
\end{equation}
where $\mathcal{D}z_\Gamma(t)$ is the integration measure over the
path $z_\Gamma(t)$, which is such that $z_\Gamma(t_a)=z_a$ and
$z_\Gamma(t_b)=z_b$. The phase factor is computed from the action
along each path given by
\begin{equation}\label{SGamma}
S_\Gamma(z_b,t_b;z_a,t_a) = \int_{t_a}^{t_b} \ud t \,L
\left[z_\Gamma(t),\dot{z}_\Gamma(t)\right]\,.
\end{equation}
It can be shown \cite{Storey} that in the case where the Lagrangian is
at most \textit{quadratic} in positions $z$ and velocities $\dot{z}$,
i.e. is of the general type
\begin{equation}\label{quadratic}
L\left[z,\dot{z}\right] =
a(t)\,\dot{z}^2+b(t)\,\dot{z}z+c(t)\,z^2+d(t)\,\dot{z}+e(t)\,z+f(t)\,,
\end{equation}
where $a(t)$, $b(t)$, $c(t)$, $d(t)$, $e(t)$ and $f(t)$ denote some
arbitrary functions of time $t$, the propagator takes the simple form
\begin{equation}\label{propquad}
K(z_b,t_b;z_a,t_a) =
F(t_b,t_a)\,\text{e}^{\frac{\rm{i}}{\hbar}
S_\text{cl}(z_b,t_b;z_a,t_a)}\,.
\end{equation}
Here the function $F$ depends on times $t_a$, $t_b$ but not on
positions $z_a$, $z_b$, which is no longer the case when the
Lagrangian is more than quadratic \cite{Storey}. We denote by
$S_\text{cl}$ the \textit{classical} action
\begin{equation}\label{Scl}
S_\text{cl}(z_b,t_b;z_a,t_a) = \int_{t_a}^{t_b} \ud t
\,L\left[z_\text{cl},\dot{z}_\text{cl}\right]\,,
\end{equation}
where the integral extends over the classical path $z_\text{cl}(t)$
satisfying the Lagrange equation
\begin{equation}\label{lagrange}
\frac{\ud}{\ud t}\left(\frac{\partial L}{\partial \dot{z}_\text{cl}}\right)
= \frac{\partial L}{\partial z_\text{cl}}\,,
\end{equation}
with boundary conditions $z_\text{cl}(t_a)=z_a$ and
$z_\text{cl}(t_b)=z_b$. One can show that the wave function deduced
from \eqref{fonctiononde} and \eqref{propquad} obeys the Schr\"odinger
equation corresponding to the Hamiltonian $H$ deduced from the
Lagrangian $L$.

Next the result \eqref{propquad} is used to
obtain the final wavefunction \eqref{fonctiononde} in the case where
the initial state $\psi(z_a,t_a)$ is a plane wave, i.e.
\begin{equation}\label{planewave}
  \psi(z_a,t_a)=\frac{1}{\sqrt{2\pi\hbar}}\,\text{e}^{\frac{\rm{i}}{\hbar}
    \left(p_0 z_a-E_0 t_a\right)}\,.
\end{equation}
Denoting by $z_0$ the initial position at time $t_a$ such that the
initial momentum of the classical path starting from $(z_0,t_a)$
equals the momentum
$p_0$ of the initial plane wave, we obtain
\begin{equation}\label{finalstate}
\psi(z_b,t_b) = G(t_b,t_a)\,\psi(z_0,t_a)\,\text{e}^{
\frac{\rm{i}}{\hbar}S_\text{cl}(z_b,t_b;z_0,t_a)}\,.
\end{equation}
The function $G$ is another function depending only on times $t_a$ and
$t_b$. The result \eqref{finalstate} is at the basis of the
computation of the phase shift in atom interferometers. We shall from
now on change our notation $z_0$ back into $z_a$ so that
\eqref{finalstate} applies to say the half upper path between
$(z_a,t_a)$ and $(z_b,t_b)$ of the interferometer depicted in
Fig.~\ref{fig1}. We emphasize at this point that this formula is valid
only for quadratic Lagrangians of the form \eqref{quadratic}. For more
general Lagrangians, the function $G$ a priori depends on the
positions $z_a$ and $z_b$, and the simple formula \eqref{finalstate}
is a priori incorrect.

Evidently, any attempt like the one proposed by MPC in
\cite{Mueller} to integrate the classical action \eqref{Scl} over
a path which is different from the classical path, i.e. which does
not obey the Lagrange equation \eqref{lagrange} or is solution of
a Lagrange equation with a different Lagrangian, will violate the
fundamental Feynman path integral formulation of quantum
mechanics. The wave function \eqref{finalstate} is then no longer
a solution of the Schr\"odinger equation and it becomes
inconsistent to use it for the calculation of the phase
difference. This inconsistency is inherent to the class of
formalisms investigated in Sec.~\ref{IIIA}, and, as we shall see,
is the only way to substantiate the claim made by MPC.

\subsection{Computation of the phase shift}\label{IIB}

\begin{figure}
\begin{center}
\includegraphics[width=13cm,angle=0]{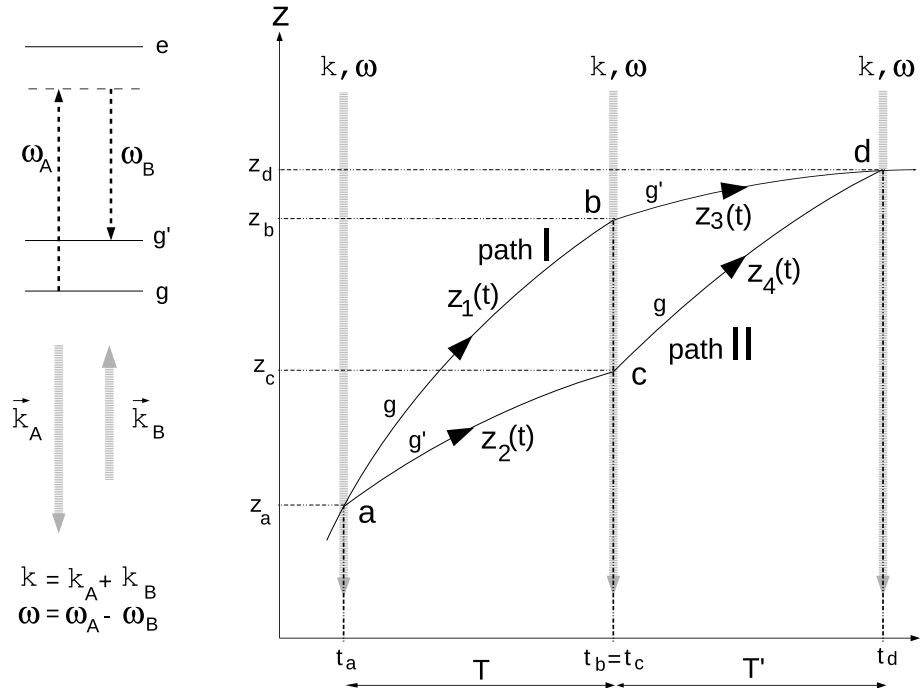}
\caption{Space-time trajectories followed by the atoms in the
  interferometer. Laser pulses occur at times $t_a$, $t_b=t_c$ and
    $t_d$, separated by time intervals $T$ and $T'$. The two-photon
    Raman transitions are between two hyperfine levels $g$ and $g'$ of
    the ground state of an alkali atom.}
\label{fig1}
\end{center}
\end{figure}

The atomic gravimeter is described in detail in
\cite{Borde89,Kasevich,Peters,Wolf} and the general theory of such
gravito-inertial devices can be found, for example, in
\cite{Storey,Borde01,Borde02}. The schematic view of the atom
interferometer showing the two interferometer paths I and II is given
by Fig.~\ref{fig1}. Here we recall only the main points relevant to
our discussion (see the detailed description of the experimental setup
in Ref.~\cite{Peters}).

The C{\ae}sium (or some alkali) atoms are optically cooled and
launched in a vertical fountain geometry. They are prepared in a
hyperfine ground state $g$. A sequence of vertical laser pulses
resonant with a $g \rightarrow g'$ hyperfine transition is applied to
the atoms during their ballistic (i.e. free fall) flight. In the
actual experiments the atoms undergo a two-photon Raman transition
where the two Raman laser beams are counter-propagating. This results
in a recoil velocity of the atoms, with the effective wave vector
transferred to the atoms being $k=k_A+k_B$, where $A$ and $B$ refer to
the counter-propagating lasers (see e.g. \cite{Storey,Wolf}). A first
pulse at time $t_a$ splits the atoms into a coherent superposition of
hyperfine states $gg'$ with the photon recoil velocity yielding a
spatial separation of the two wave packets. A time interval $T$ later
(we denote $T=t_b-t_a=t_c-t_a$; see Fig.~\ref{fig1}) the two wave
packets are redirected toward each other by a second laser pulse
thereby exchanging the internal states $g$ and $g'$. Finally a time
interval $T'$ later (with $T'=t_d-t_b=t_d-t_c$) the atomic beams
recombine and a third pulse is applied. After this pulse the
interference pattern in the ground and excited states is measured.

The calculation of the phase shift $\Delta \varphi$ of the atomic
interferometer in the presence of a gravitational field was done in
Refs.~\cite{Borde89,Kasevich,Storey} and proceeds in several
steps. The first contribution to the phase shift comes from the free
propagation of the atoms in the two paths. Using the result
\eqref{finalstate} of the path integral formalism, and the fact that
the function $G$ therein depends on times $t_a,t_b$ but not on
positions $z_a,z_b$, we see that this contribution is equal to the
difference of classical actions in the two paths,
\begin{equation}\label{action_phase}
\Delta \varphi_S = \frac{\Delta S_\text{cl}}{\hbar}\,.
\end{equation}
To compute it one calculates the classical trajectories of the wave
packets in the two arms using the equations of motion of massive test
bodies deduced from the classical Lagrangian and the known boundary
conditions (position and momenta) of the wave packets. In the case of
the general quadratic Lagrangian \eqref{quadratic} the equations of
motion read
\begin{equation}\label{eom}
\frac{\ud}{\ud t}\Bigl[2a(t) \dot{z}\Bigr] = \bigl[2c(t) -
\dot{b}(t)\bigr] z + e(t) -\dot{d}(t)\,.
\end{equation}
Following Fig.~\ref{fig1} we denote by $z_1(t)$ and $z_3(t)$ the
classical trajectories between the laser interactions in the upper
path I, and by $z_2(t)$ and $z_4(t)$ the trajectories in the lower
path II. The boundary conditions in positions appropriate to the
closed-path interferometer of Fig.~\ref{fig1} are (for simplicity we
set $t_a=0$)
\begin{subequations}\label{cont}{\begin{align}
& z_1(0) = z_2(0)\,,\\
& z_1(T) = z_3(T)\,,\\
& z_2(T) = z_4(T)\,,\\
& z_3(T+T') = z_4(T+T')\,.
\end{align}}\end{subequations}
The boundary conditions in velocities are determined by the recoils
induced from the interactions with the lasers. Actually once we have
imposed the boundary conditions \eqref{cont}, and in particular that
the interferometer closes at some time $T+T'$, only the recoils due to
the second pulse at the intermediate time $T$ are needed for this
calculation. These are given by
\begin{subequations}\label{deltav}{\begin{align}
\dot{z}_1(T)-\dot{z}_3(T)&=+\frac{\hbar k}{m}\,,\\
\dot{z}_2(T)-\dot{z}_4(T)&=-\frac{\hbar k}{m}\,,
\end{align}}\end{subequations}
where $k=k_A+k_B$ is the effective wave vector transferred by the
lasers to the atoms. Second, one calculates the difference in the
classical action integrals along the two paths,
\begin{equation}\label{DS}
  \Delta S_\text{cl} = \int_0^T \Bigl(L[z_1,\dot{z}_1]-L[z_2,\dot{z}_2]\Bigr)
  \ud t
  + \int_T^{T+T'} \Bigl(L[z_3,\dot{z}_3]-L[z_4,\dot{z}_4]\Bigr)\ud t
  + \Delta S_{gg'}\,,
\end{equation}
where the Lagrangian $L(z,\dot{z})$ is given by \eqref{quadratic}, and
the integrals are carried out along the classical paths calculated in
the first step. We have taken into account the changes in energy
between the hyperfine ground states $g$ and $g'$ of the atoms in each
path. These energies will cancel out from the two paths provided that
$T'$ is equal to $T$, which will be true for a Lagrangian in which we
neglect gravity gradients. In the more general case $T'$ will differ
from $T$ and there is an extra contribution in \eqref{DS} given by
\begin{equation}\label{DeltaSgg'}
\Delta S_{gg'} = \hbar \,\omega_{gg'}(T-T')\,,
\end{equation}
where we denote the internal energy change by
$E_{g'}-E_{g}=\hbar\,\omega_{gg'}$.

We now prove (see e.g. \cite{Storey,Borde01}) that the two action
integrals in \eqref{DS} cancel each other in the case of a quadratic
Lagrangian of the general form \eqref{quadratic}. This follows from
the fact that the difference between the Lagrangians
$L[z_1(t),\dot{z}_1(t)]$ and $L[z_2(t),\dot{z}_2(t)]$, which are
evaluated at the same time $t$ but on two different trajectories
$z_1(t)$ and $z_2(t)$, is a total time-derivative when the Lagrangians
are ``on-shell'', i.e. when the two trajectories $z_1(t)$ and $z_2(t)$
satisfy the equations of motion \eqref{eom}.  Namely, we
find\footnote{To prove this we consider the first contribution
  $a(\dot{z}^2_1-\dot{z}^2_2)$ in the difference $L_1-L_2$ and
  re-express it thanks to an integration by parts as
  $a(\dot{z}^2_1-\dot{z}^2_2) = \frac{\ud}{\ud
    t}[a(z_1-z_2)(\dot{z}_1+\dot{z}_2)]-(z_1-z_2)\frac{\ud}{\ud
    t}[a(\dot{z}_1+\dot{z}_2)]$. The second term is then simplified by
  means of the sum of the equations of motion \eqref{eom} written for
  $z=z_1$ and $z=z_2$. In addition we also integrate by parts the
  second and fourth contributions to $L_1-L_2$ as
  $b(\dot{z}_1z_1-\dot{z}_2z_2)=\frac{\ud}{\ud
    t}[\frac{1}{2}b(z^2_1-z^2_2)]-\frac{1}{2}\dot{b}(z^2_1-z^2_2)$ and
  $d(\dot{z}_1-\dot{z}_2)=\frac{\ud}{\ud
    t}[d(z_1-z_2)]-\dot{d}(z_1-z_2)$. Summing up the results we obtain
  \eqref{totaltimeder}.}
\begin{equation}\label{totaltimeder}
L[z_1,\dot{z}_1] - L[z_2,\dot{z}_2] = \frac{\ud}{\ud
t}\biggl[(z_1-z_2)\biggl(a(t)(\dot{z}_1+\dot{z}_2)+\frac{1}{2}b(t)(z_1+z_2)
+d(t)\biggr)\biggr]\,.
\end{equation}
Since the difference of Lagrangians is a total time derivative the
difference of action functionals in \eqref{DS} can be immediately
integrated. Using the continuity conditions at the interaction points
with the lasers \eqref{cont}, and then using the recoils due to the
intermediate laser pulse at time $T$ given by \eqref{deltav}, yields
\begin{equation}\label{DSres}
\Delta S_\text{cl} - \Delta S_{gg'} =
a(T)\bigl[z_1(T)-z_2(T)\bigr]\Bigl[\dot{z}_1(T)
+\dot{z}_2(T)-\dot{z}_3(T)-\dot{z}_4(T)\Bigr] = 0\,.
\end{equation}
Therefore for all quadratic Lagrangians the difference of classical
actions in the interferometer, and therefore the phase shift due to
the free propagation of the atoms, reduces to the contribution of the
change of internal states $g$ and $g'$, thus\footnote{Rigorously, in
  this equation the time interval should be a proper time interval.}

\begin{equation}\label{Dphiresult}
\Delta \varphi_S = \omega_{gg'}(T-T')\,.
\end{equation}
When the interferometer is symmetric, i.e. $T=T'$ which will be the
case when we neglect gravity gradients, we get exactly zero. Equations
\eqref{DSres} or \eqref{Dphiresult} constitute the central theorem to
be used for the discussion in the present paper.

In the second step, one calculates the difference in the light phases
$\Delta\varphi_\ell$ at the three interactions with lasers. These are
obtained using the paths calculated in the first step and the
equations of light propagation, with the light acting as a ``ruler''
that measures the motion of the atoms. For the interferometer
described in Fig.~\ref{fig1} the phase difference from light
interactions is given by
\begin{equation}\label{phaselight}
\Delta \varphi_\ell = - \phi(z_a,0) + \phi(z_b,T)
+ \phi(z_c,T) - \phi(z_d,T+T')\,.
\end{equation}
Here $\phi$ denotes the phase of the laser light as seen by the atom,
i.e. $\phi(z,t) = k z - \omega t - \phi_0$ where $k$, $\omega$ and
$\phi_0$ are the wave vector, the frequency and the phase of the laser
in the frame of the laboratory. The phase of the laser is evaluated at
the interaction points shown in Fig.~\ref{fig1}. Finally the total
phase shift measured in the atom interferometer is the sum of the two
contributions
\begin{equation}\label{DeltaAll}
\Delta\varphi = \omega_{gg'}(T-T') + \Delta \varphi_\ell\,,
\end{equation}
depending only on the internal states $g$ and $g'$, and the light
phases which measure the free fall trajectories of the atoms.

Here we have assumed that the two paths I and II close up at the entry
and exit of the interferometer (see Fig.~\ref{fig1}). For the more
general case where the paths are not closed, one finds that $\Delta
\varphi_S$ is no longer zero, but that $\Delta \varphi_S$ is exactly
cancelled by additional terms corresponding to the phase difference of
the wave packets at different positions at the interferometer entry
and exit \cite{Borde01} (except for the very small terms due to the
energy difference $\hbar\omega_{gg'}$).

Expression \eqref{Dphiresult} shows that, depending on the particular
geometry of the interferometer, one can obtain a sensitivity to the
frequency difference $\omega_{gg'}$ of the two internal states. In
that sense the interferometer can be viewed as a clock. In fact,
atomic clocks are a particular kind of interferometer which has been
designed to have a strong sensitivity to $\omega_{gg'}$
\cite{Borde01,Borde02}. A simple example is an interferometer with
only two pulses separated by a time $T$. In that case one uses either
a microwave field or Raman transitions with two co-propagating laser
beams so that $k=k_A-k_B$. Then the photon recoil becomes negligibly
small and the two paths are separated spatially at $t=T$ by much less
than the coherence length of the atoms, allowing them to
interfere. The total phase shift is then given by
\begin{equation}\label{DeltaClock}
\Delta\varphi = \omega_{gg'}T + \Delta \varphi_\ell\,.
\end{equation}
As mentioned above, all other terms in $\Delta\varphi_S$ are exactly
cancelled by corresponding terms arising from the non-closure of the
paths. The laser phase shift is to leading order
\begin{equation}\label{phaselightClock}
\Delta \varphi_\ell = - \omega T\,.
\end{equation}
One therefore observes interference fringes when varying the frequency
of the laser or microwave field, known as Ramsey fringes. That allows
locking the laser frequency to the atomic transition frequency
$\omega_{gg'}$, which constitutes an atomic clock. We note that the
\textit{interference} of the two interferometer arms is required for
the operation of the clock, each arm individually cannot be considered
as a clock.

If one wanted to use such interferometers (clocks) for a
gravitational redshift test one would need two of them placed in
different gravitational potentials (e.g.
\cite{Vessot,Cacciapuoti}). But, of course such clocks do not run
at the Compton frequency but at $\omega_{gg'}$, and thus the
corresponding redshift test is not at the Compton frequency. For
that to be the case one would need interferometers with the
internal energy (mass) \textit{difference} of the particles on the
two paths $\hbar \omega_{gg'}=m c^2$. Such a matter-antimatter
interferometer is orders of magnitude beyond present day
technology.

\subsection{Prediction from general relativity}\label{IIC}

As a specific example, let us consider the prediction from GR, which
has been extensively treated in Ref.~\cite{Borde01}; here we only
present a very basic analysis sufficient for our purposes. The
appropriate Lagrangian is derived to sufficient accuracy using
the Schwarzschild metric generated by the Earth,
\begin{equation}
  L_\text{GR}(z,\dot{z})= -mc^2\frac{\ud\tau}{\ud t}=-mc^2+\frac{GMm}{r_\oplus}
  -mgz+\frac{1}{2}m\dot{z}^2
  +\mathcal{O}\left(\frac{1}{c^2}\right)\,, \label{LGR}
\end{equation}
where $\ud\tau=\sqrt{-g_{\mu\nu}\ud x^\mu \ud x^\nu/c^2}$ is the
proper time, $r_\oplus$ is the Earth's radius, $g={GM}/{r_\oplus^2}$
is the Newtonian gravitational acceleration, $G$ is Newton's
gravitational constant, $M$ is the mass of the Earth, $m$ the mass of
the atom, $c$ the speed of light in vacuum, $z$ is defined by
$r=r_\oplus+z$ with $r$ the radial coordinate, and
$\mathcal{O}(1/c^2)$ denotes a post-Newtonian correction. For
simplicity we restrict ourselves to only radial motion, which is
sufficient for the arguments presented in this paper. We also neglect
systematically the post-Newtonian correction and no longer indicate
the remainder term $\mathcal{O}(1/c^2)$.

The equations of motion are deduced from \eqref{LGR} using the
principle of least action or the Euler-Lagrange equations, and read
evidently $\ddot{z}=-g$. Then, when integrating \eqref{LGR} along the
resulting paths, and calculating the difference as defined by
\eqref{DS} one finds
\begin{equation}\label{DeltaphiS}
\Delta \varphi_S = 0\,.
\end{equation}
Since the Lagrangian \eqref{LGR} is quadratic, this is a particular
case of the general result \eqref{DSres}. Furthermore, in the simple
case of a Lagrangian which is linear in $z$ we find that the
interferometer is symmetrical, i.e. the interrogation times are equal,
$T=T'$.  Thus the effect due to the difference of internal energies
$E_{g'}-E_{g}$ is zero in this case.

Evaluating the light phases at the different interaction points one
finds easily that (see \cite{Storey,Borde01,Peters,Wolf} for details)
\begin{equation}\label{Deltaphiell}
\Delta \varphi_\ell = k\,g\,T^2\,,
\end{equation}
where $k$ is the effective wave vector of the lasers (i.e.
$k=k_A+k_B$ where the subscripts A and B refer to the
counter-propagating laser beams \cite{Wolf}), and $T$ is the time
interval between the laser pulses (see Fig.~\ref{fig1}). As a result
the total phase shift calculated in GR is
\begin{equation}\label{Deltaphitotal}
\Delta \varphi = k\,g\,T^2\,.
\end{equation}
This shows that the atom interferometer is a gravimeter or
accelerometer. The phase shift \eqref{Deltaphitotal} arises entirely
from the interactions with the lasers and the fact that the atoms are
falling with respect to the laboratory in which the experiment is
performed. It is thus proportional to the acceleration $g$ of atoms
with respect to the experimental platform which holds the optical and
laser elements.  With $k$ and $T$ known from auxiliary measurements,
one deduces the component of $g$ along the direction of $k$. If the
whole instrument was put into a freely falling laboratory, the
measured signal $\Delta\varphi$ would vanish.\footnote{The result
    is identical for the phase shift in neutron interferometers under
    the influence of gravity, as shown in Appendix \ref{appA}.}

The signal \eqref{Deltaphitotal} is actually the same as in the
measurement of the free fall of a macroscopic object (corner cube)
using lasers. Therefore atomic interferometers can be used for testing
the universality of free fall between the atoms and some macroscopic
objects. The precision on the test of the UFF between C{\ae}sium atoms
and classical objects such as the corner cube is currently
$7\times10^{-9}$ \cite{Peters,Mueller,refsyrte}.

We have stressed that integrating the Lagrangian along the classical
paths only provides the correct phase shift when the Lagrangian is at
most quadratic in position and velocity \cite{Storey}. For higher
order Lagrangians the full path integrals need to be evaluated. Thus,
one might be tempted to use atom interferometry tests to search for
redshift effects arising from third and higher order terms in the
Lagrangian, but as already mentioned and shown explicitly in
e.g. \cite{Storey} such an approach is incorrect, as the full Feynman
integral is required.\footnote{If one nonetheless calculates the
phase shift for a higher-order Lagrangian by integrating only along
the classical path, a non zero result is obtained. In Appendix
\ref{appB} we estimate the order of magnitude of the effect of
including cubic terms in the Lagrangian of GR, which correspond to
second-order gravity gradients.}

\section{Testing the gravitational redshift\,?}\label{III}

We now address the question of the significance of tests of the
gravitational redshift by atom interferometers as described by MPC
\cite{Mueller} within the frameworks of different classes of
alternative theories or models. Depending on the underlying theory,
the Lagrangian will be different, so the equations of motion providing
the paths will be different, and the relation between the action
integral and the phase shift [cf. Eq.~\eqref{action_phase}] may also
be different. In addition we may have to model differently the
propagation of light, so the resulting calculated phase shift might
also differ. The comparison of the calculated and measured phase
shifts then allows to discriminate between different candidate
theories. We shall see that the significance of the tests depends on
the alternative model we consider.

\subsection{Multiple Lagrangian framework}\label{IIIA}

In this formalism the atom interferometry experiment could be
viewed as testing the gravitational redshift or the universality
of clock rates (UCR), and it is the formalism used by MPC in
\cite{Mueller}.
More precisely, this corresponds to the first scenario of MPC (see
  ``Methods'', bottom of first column), where they calculate the phase
  shift by integrating the Lagrangian with $g(1+\beta)$ along the
  trajectories calculated with the standard value of
  $g$.\footnote{In the second scenario, MPC still integrate the
    Lagrangian with $g(1+\beta)$ but use $g'$ different from $g$ for
    the trajectories without explaining clearly how to calculate
    $g'$.}

Within this multiple Lagrangian framework we derive the
interferometer phase shift (putting on a more rigorous basis the
computation of
  MPC) and then we show that this framework raises a number of
unacceptable issues related to fundamental principles of quantum
mechanics.

The framework is motivated by a search for a possible violation of the
local position invariance (LPI) aspect of the Einstein equivalence
principle, while supposing that the remaining aspects of the
equivalence principle, namely the local Lorentz invariance (LLI) and
the weak equivalence principle (WEP), are valid. Thus, in the first
place, this framework violates the Schiff conjecture
\cite{Schiff}. (See Ref.~\cite{Will} for a thorough discussion on the
different parts of the Einstein equivalence principle and the Schiff
conjecture.)

If WEP is valid we can consider the local freely falling frames
associated with test bodies, which fall with the universal
acceleration $g$ in a gravitational field. In these frames classical
clocks will measure a proper time $\ud\tau$ which is proportional,
because LLI is valid, to the special-relativistic Minkowskian interval
$\ud s$. However, because LPI is violated, we allow for a dependence
on the position of the measured rate. Specifically we introduce a
proportionality factor $f(\Phi)$ between $\ud \tau$ and $\ud s$ built
from some anomalous field $\Phi$ associated with gravity and depending
on position. In an arbitrary frame this means that the proper time
$\ud \tau$ measured by clocks is proportional to the metric interval
$\ud s$ and given by
\begin{equation}\label{dtau}
\ud \tau = f(\Phi) \,\ud s = f(\Phi)\sqrt{-g_{\mu\nu}\ud x^\mu \ud
x^\nu /c^2}\,.
\end{equation}
But because WEP is valid the motion of test bodies is unaffected and
obtained from the usual variational principle associated with the
metric interval $\ud s$, i.e. $\delta \int \ud s = 0$.

Consider a classical redshift experiment. One observes at point $z_1$
the light coming from a particular atomic transition occuring at point
$z_0$. The points $z_1$ and $z_0$ are at rest in a stationary
gravitational field $g$. The relative difference between the frequency
of the light coming from point $z_0$ and observed at point $z_1$, with
respect to the frequency of the light emitted by the same atomic
transition but occuring at point $z_1$, is
\begin{equation}\label{Z}
Z\equiv \frac{\Delta \tau_1}{\Delta \tau_0} - 1 =
\frac{f(\Phi_1)}{f(\Phi_0)}\frac{\sqrt{-g_{00}(z_1)}}{\sqrt{-g_{00}(z_0)}}
- 1\,,
\end{equation}
where $\Phi_1=\Phi(z_1)$ and $\Phi_0=\Phi(z_0)$. It is convenient to
introduce a coefficient $\zeta$ linking the gradient of the anomalous
field to the local gravitational acceleration by $g=\zeta\,\ud\Phi/\ud
z$ (see \cite{Will} for more details). Expanding to first order in
$\Delta z=z_1-z_0$ we can write
\begin{equation}\label{f}
f=f_0\left[ 1 +\beta \,\frac{g \Delta z}{c^2}\right]\,,
\end{equation}
where we have defined
\begin{equation}\label{beta}
\beta=\frac{c^2f'_0}{\zeta\,f_0}\,,
\end{equation}
with $f_0$ and $f'_0$
denoting $f$ and $\ud f/\ud\Phi$ evaluated at the value
$\Phi(z_0)$. Inserting then \eqref{f} into \eqref{Z} we find that
the parameter $\beta$ defined in this way measures a possible
anomalous deviation from the standard prediction for the redshift,
i.e.
\begin{equation}\label{Zbeta}
Z = \left( 1 +\beta \right) \frac{g \Delta z}{c^2}\,,
\end{equation}
and we know that $\beta$ has been tested by redshift experiments with
the accuracy $10^{-4}$ \cite{Vessot}.

Usually the parameter measuring redshift violations is denoted
$\alpha$ but here it is important to adopt the different notation
$\beta$ used by MPC. Indeed one should recall that the meaning of
parameters $\alpha$ or $\beta$ depends on the theoretical framework in
use. The result \eqref{Zbeta} corresponds to our particular formalism,
the multiple Lagrangian formalism, and should be contrasted with a
similar result \eqref{Zalpha} we shall obtain below in the context of
the modified Lagrangian formalism in Sec.~\ref{IIIB}. In particular we
notice that the parameter $\beta$ in Eq.~\eqref{Zbeta} may or may not
be ``universal'', i.e.  depending on the nature and internal structure
of the clock, in contrast with the parameter $\alpha^{(a)}_X$ in
Eq.~\eqref{Zalpha} which depends on the type of atom $(a)$
constituting the clock and on some type of internal energy $X$
violating LPI. In the present formalism we do not need to specify if
$\beta$ is universal or not because WEP is valid so we can always test
the value of $\beta$ by measuring $g$ from the free-fall of test
bodies (as proposed by MPC). This is different in the modified
Lagrangian formalism of Sec.~\ref{IIIB} where any universal
contribution to $\alpha^{(a)}_X$ is unobservable since it can always
be absorbed into a re-definition of $g$.

Let us now apply this formalism to atom interferometry, viewing the
atoms in the two paths of the interferometer as classical clocks
ticking at the Compton frequency $\omega_\text{C}$. In this case $z_1$
is a generic position of the atom on one of the paths (see
Fig.~\ref{fig1}) and $z_0$ denotes an origin located between the
paths. Following the analogy with classical clocks discussed in
Sec.~\ref{IB} of the Introduction, we can write the phase difference
between the two ``atom-clocks'' for a closed interferometer
as\footnote{As is clear from e.g. \eqref{dtau}, the constant $f_0$
    can be absorbed into a rescaling of the coordinates by posing
    $x_\text{new}^\mu=f_0\,x^\mu_\text{old}$. Henceforth we set this
    constant to $f_0=1$.}
\begin{equation}\label{Dphimueller0}
\Delta\varphi_S = \omega_\text{C} \oint \ud \tau = \omega_\text{C}
\oint \left[ 1 +\beta \,\frac{g \Delta z}{c^2}\right]\ud s \,.
\end{equation}
On the other hand the trajectories of the atoms are the usual
geodesics of space-time, such that $\delta \int \ud s = 0$. As we have
proved in Sec.~\ref{IIB}, in the case of a quadratic Lagrangian and a
symmetric closed interferometer, the integral of $\ud s$ is zero, so
one is left with
\begin{equation}\label{Dphimueller1}
\Delta\varphi_S = \omega_\text{C} \oint \frac{\beta g z}{c^2} \ud s \,.
\end{equation}
At the dominant order we can approximate $\ud s$ by the coordinate
time $\ud t$ and a short calculation shows that the phase shift is non
zero but given by
\begin{equation}\label{Dphimueller2}
\Delta\varphi_S = \beta \,k \,g \,T^2\,.
\end{equation}
Adding the term due to the interaction with lasers which is standard
(because WEP is valid) and given by \eqref{Deltaphiell}, we obtain the
total phase shift
\begin{equation}\label{Dphimueller}
\Delta\varphi = \Delta\varphi_S + \Delta\varphi_\ell
= \left( 1 +\beta \right) k \,g \,T^2\,,
\end{equation}
where $\beta$ takes the same meaning as the redshift violation
parameter entering \eqref{Zbeta}.

Comparing the predictions \eqref{Zbeta} and \eqref{Dphimueller} we
conclude that within this formalism atom-interferometric gravity
measurements would test the gravitational redshift, and would improve
existing limits on this measurement by four orders of magnitude, as
claimed by MPC.

However, let us now critically examine the implications and meaning of
this formalism, and show that it raises a number of
difficult fundamental problems.  We refer to this formalism (or class
of alternative theories) as a multiple Lagrangian framework because
for these theories the trajectories of massive bodies (classical
particles or classical paths of wave packets) obey a Lagrangian which
is different from the Lagrangian used to compute the proper time of
clocks or the phase shift of matter waves. Since WEP is valid the
motion of massive classical particles obeys the standard Lagrangian of
general relativity, $L_\text{particle} = L_\text{GR} = -mc^2\ud s/\ud
t$, i.e. to first order
\begin{equation}\label{Lpart}
L_\text{particle} = -m\,c^2+\frac{GMm}{r_\oplus}-m\,g\,z+\frac{1}{2}m\dot{z}^2\,.
\end{equation}
On the other hand the action integral we have used for the computation
of the phase shift of the quantum matter wave is calculated from a
different Lagrangian given by
\begin{equation}\label{Lwave}
L_\text{wave} = \left(1+\beta\,\frac{g\,z}{c^2}\right)L_\text{particle}\,.
\end{equation}
We first note that this Lagrangian is not quadratic, and,
  neglecting higher-order corrections, we recover the Lagrangian used
  by MPC \cite{Mueller}, i.e.
\begin{equation}\label{Lmueller}
L_\text{wave} =-m\,c^2+\frac{GMm}{r_\oplus}-(1+\beta)m\,g\,z
+\frac{1}{2}m \dot{z}^2\,,
\end{equation}
where the parameter $\beta$ that measures the deviation from GR
(i.e. $\beta=0$ in GR) enters as a correction in the atom's
gravitational potential energy.

The most important problem is that it is inconsistent to use a
different Lagrangian (or metric) for the calculation of the
trajectories and for the phases of the atoms or clocks. More
precisely, it supposes that the fundamental Feynman path integral
formulation of quantum mechanics, which is at the basis of the
derivation of the phase shift in an atom interferometer (see
Sec.~\ref{IIA}), has to be altered in the presence of a gravitational
field or could be wrong.\footnote{In the case of atom gravimeters
  using Bloch oscillation \cite{Clade,Ivanov}, also interpreted as
  redshift tests by MPC, the atoms are ``trapped'' in optical dipole
  traps, and the Bloch oscillation frequencies are interpreted as
  interference of two matter waves at Compton frequency, without using
  the Feynman formalism. However, this interpretation is plagued with
  the same difficulties. The Bloch oscillation frequency is obtained
  from the Schr\"odinger equation with the Hamiltonian including a
  gravitational term. We first note that adding or not the rest-mass
  energy to that Hamiltonian does not modify the derived Bloch
  frequency, hence the Compton frequency of the atoms plays no
  role. Second, to obtain the result of MPC one needs to use a
  different gravitational term in the Hamiltonian when calculating the
  states and energies providing the Bloch frequency or when
  calculating the motion of the atoms in the same gravitational
  field. This is akin to the multiple Lagrangian framework with the
  same difficulties and inconsistencies.} Physically it amounts to
making the distinction between the atoms when calculating their
trajectories and the same atoms when calculating their phases, which
is inconsistent as it is the same fundamental matter field in both
cases. Even more, the basis for writing Eq.~\eqref{Dphimueller0} for
the derivation of the atom interferometry phase shifts (as does MPC)
is unjustified because the Feynman formalism is violated. To remain
coherent some alternative formalism for the atomic phase shift
calculations (presumably modifying quantum mechanics) should be
developed and used. More generally the multiple Lagrangian formalism
supposes that the duality between particles and waves in quantum
mechanics gets somehow violated in a gravitational field.

The above violation of the ``particle-wave duality'', implied by the
multiple Lagrangian formalism, is very different from a violation of
the equivalence principle in the ordinary sense.  In this formalism a
single physical object, the atom, is assumed to be described by two
different Lagrangians, $L_\text{wave}$ applying to its phase shift,
and $L_\text{particle}$ applying to its trajectory. By contrast, in
tests of the equivalence principle, one looks for the modification of
the free fall trajectories or clock rates as a function of composition
or clock type. Thus, different bodies or clocks are described by
different Lagrangians, but of course for any single type of body we
always have $L_\text{wave}=L_\text{particle}$. The pertinent test-bed
for equivalence principle violations is the modified Lagrangian
formalism reviewed in Sec.~\ref{IIIB}, which does not imply any
particle-wave duality violation.

The second problem of theories in the multiple Lagrangian formalism
and of the interpretation of MPC is the violation of Schiff's
conjecture \cite{Schiff}. Indeed, we have assumed in this Section that
the LPI aspect of the equivalence principle is violated but that for
instance the WEP aspect remains satisfied. In a complete and
self-consistent theory of gravitation one expects that the three
aspects of the Einstein equivalence principle (WEP, LLI and LPI) are
sufficiently entangled together by the mathematical formalism of the
theory that it is impossible to violate one without violating all of
them. The Schiff conjecture has been proved using general arguments
based upon the assumption of energy conservation
\cite{Nordtvedt,Haugan}. A contrario, we expect that violating the
conjecture leads to some breakdown of energy conservation. In the
present case this translates into postulating two Lagrangians for the
same physical object. Explicit theories that are logically and
mathematically consistent but still violate Schiff's conjecture are
very uncommon (see \cite{Will} and references therein).\footnote{Ni
  \cite{Nischiff} has proposed a possible counterexample of Schiff's
  conjecture.  As it is based on a modified Lagrangian, it falls in
  the category of theories investigated in Sec.~\ref{IIIB}, for which
  atom interferometry does not test the redshift.}

In conclusion the multiple Lagrangian formalism violates the Feynman
path integral formulation of quantum mechanics, the principle of least
action and energy conservation, and Schiff's
conjecture. Unfortunately, the claim made by MPC \cite{Mueller} that
atom gravimeters measure the gravitational redshift can only be
substantiated within this formalism.

\subsection{Modified Lagrangian framework}\label{IIIB}

In this Section we review and adapt for our purpose the important
``modified Lagrangian framework'', which is a powerful formalism
for analyzing tests of the various facets of the Einstein
equivalence principle (EEP). This formalism allows deviations from
GR and metric theories of gravity, with violations of the UFF and
UCR, and permits a coherent analysis of atom interferometry
experiments. But it does not suffer from the shortcomings of the
multiple Lagrangian framework. The formalism describes the
Newtonian limit of a large class of non-metric theories, in a way
consistent with Schiff's conjecture and fundamental principles of
quantum mechanics.

Within this general
  formalism there is no fundamental distinction between UFF and UCR
  tests, as violation of one implies violation of the other (Schiff's
  conjecture), thus testing one implies testing the other. Depending
  on the theory used, different experiments test different parameters
  or parameter combinations at differing accuracies, so they may be
  complementary or redundant depending on the context. As we shall
  see, in such theories the mass of the atoms cancels out when
  computing the phase shift of the atom interferometer, and thus the
  atom's Compton frequency is irrelevant. In that sense atom
  interferometers test UFF (and are indeed interpreted that way in the
  more recent work \cite{Hohensee2}), but are insensitive to a test of
  UCR at the Compton frequency.

This class of theories is defined by a single Lagrangian, that is
however different from the GR Lagrangian \eqref{LGR}. In
particular, the coupling between gravitation and different types
of mass-energy is generically not universal \cite{Damour}. This
leads to modifications of the gravitational redshift and also,
more generally, to violations of the weak equivalence principle or
universality of free fall, the local Lorentz invariance and the
local position invariance. But it does not imply a violation of
the principle of least action nor of energy conservation, as it is
based on a single Lagrangian and, of particular interest here, the
Feynman path integral formulation of quantum mechanics remains
valid. Most alternative theories commonly considered belong to
this class which encompasses a large number of models and
frameworks (see e.g. \cite{Will,Damour} and references therein).
It includes for example most non-metric theories (e.g. the
Belinfante-Swihart theory \cite{BelinfanteS}), some models
motivated by string theory \cite{DamourP} and brane scenarios,
some general parameterized frameworks such as the energy
conservation formalism \cite{Nordtvedt,Haugan} (see also
\cite{Dicke64}), the $\text{TH}\varepsilon\mu$ formalism and its
variants \cite{LightLee,Blanchet}, and the Lorentz violating
standard model extension (SME) \cite{Koste1,Koste2} used in
\cite{Hohensee2}.

The modified Lagrangian formalism is a physical framework in which the
violation of EEP originates from the anomalous behaviour of some
particular type of energy in a gravitational field. For our purposes
it is sufficient to use a strongly simplified ``toy'' Lagrangian
chosen as a particular case within the ``energy conservation
formalism'' of Nordtvedt and Haugan \cite{Nordtvedt,Haugan} (see
\cite{Will} for a review). To keep in line with MPC we choose an
expression similar to the Lagrangian of GR given by \eqref{LGR},
namely
\begin{equation}\label{Lmodified0}
L_\text{modified} =
-m(z)\,c^2+\frac{GMm(z)}{r_\oplus}-m(z)\,g\,z+\frac{1}{2}m(z)\,\dot{z}^2\,.
\end{equation}
We postulate that for the body or atom under consideration the mass
$m$ depends on the position $z$ through a violation of the LPI
subprinciple of EEP. This is modelled by assuming that a particular
internal energy of the atom $E_X$ behaves anomalously in the presence
of the gravitational field, where $X$ refers to the type of
interaction involved (electromagnetic, nuclear, spin-spin, spin-orbit,
etc). In the general case $E_X$ could depend on both the position $z$
and velocity $\dot{z}$ of the atom \cite{Haugan}, but here we consider
only a dependence on $z$ to model the violation of the LPI. Separating
out $E_X(z)$ from the other forms of energies $\overline{E}_Y$
composing the atom and which are supposed to behave normally, we write
\begin{equation}\label{mz}
m(z) = \overline{m} + \frac{1}{c^2}\biggl[E_X(z) + \sum_{Y\not=X}
\overline{E}_Y\biggr]\,.
\end{equation}
Here $\overline{m}$ denotes the sum of the rest masses of the
particles constituting the atom, and $\overline{m}$ and all
$\overline{E}_Y$'s are constant. The violation of LPI is modelled in
the simplest way by assuming that at the leading order
\begin{equation}\label{EX}
E_X(z) = \overline{E}_X + \beta^{(a)}_X \,\overline{m} \,g \,z\,,
\end{equation}
where $\beta^{(a)}_X$ denotes a dimensionless parameter characterizing
the violation of LPI and depending on the particular type of
mass-energy or interaction under consideration, e.g. $\beta^{(a)}_X$
would be different for the electromagnetic or the nuclear
interactions, with possible variations as a function of spin or the
other internal properties of the atom, here labelled by the
superscript $(a)$. Thus $\beta^{(a)}_X$ would depend not only on the
type of internal energy $X$ but also on the type of atom
$(a)$. Defining now the ``normal'' contribution to the total mass,
\begin{equation}
m_0 = \overline{m} + \sum_Y \frac{\overline{E}_Y}{c^2}\,,
\label{mbar}
\end{equation}
and replacing $m(z)$ by its explicit expression into the Lagrangian
\eqref{Lmodified0} we obtain
\begin{equation}\label{Lmodified}
L_\text{modified} = - m_0 \,c^2+\frac{GM m_0}{r_\oplus} -
\bigl(1+\beta^{(a)}_X\bigr) \,m_0 \,g\,z +\frac{1}{2}m_0\,\dot{z}^2\,,
\end{equation}
in which we have neglected some higher-order relativistic terms. A
major difference with the Lagrangian \eqref{Lmueller} of the previous
Section is that the coefficient $\beta^{(a)}_X$ is not universal but
depends on the internal structure of the atom.

Before discussing the atom interferometry experiment, let us recall
the results for traditional free fall and redshift experiments. By
varying \eqref{Lmodified} we obtain the equations of motion of the
atom as
\begin{equation}\label{acc}
\ddot{z} = - \bigl(1+\beta^{(a)}_X\bigr)\,g\,,
\end{equation}
which shows that the trajectory of the atom is affected by the
violation of LPI and is not universal. In fact we see that
$\beta^{(a)}_X$ measures the non-universality of the ratio between the
atom's passive gravitational mass and inertial mass. Thus, in the
modified Lagrangian framework the violation of LPI implies a violation
of WEP and the UFF, and $\beta^{(a)}_X$ appears to be the
UFF-violating parameter. This is a classic example
\cite{Nordtvedt,Haugan} of the validity of Schiff's conjecture, namely
that it is impossible to violate LPI (or LLI) without also violating
WEP.

The violation of LPI is best reflected in classical redshift
experiments with clocks which can be analysed using a cyclic gedanken
experiment based on energy conservation
\cite{Nordtvedt,Haugan,Will}. The result for the frequency shift in a
Pound-Rebka type experiment is
\begin{equation}\label{Zalpha}
Z = \bigl(1+\alpha^{(a)}_X\bigr)\,\frac{g \,\Delta z}{c^2}\,,
\end{equation}
where the redshift violating (or UCR violating) parameter
$\alpha^{(a)}_X$ is again non-universal. As such it has to be
carefully distinguished from the redshift violation parameter
$\beta$ appearing in Eq.~\eqref{Zbeta} of the previous Section
[see also the discussion after \eqref{Zbeta}]. The important
point, proven in Refs.~\cite{Nordtvedt,Haugan,Will}, is that
within the framework of the modified Lagrangian \eqref{Lmodified},
the UCR violating parameter $\alpha^{(a)}_X$ is related in a
precise way to the UFF-violating parameter $\beta^{(a)}_X$, namely
(see e.g. Eqs.~(6) and (7) in~\cite{Nordtvedt})
\begin{equation}\label{alphabeta}
\beta^{(a)}_X = \alpha^{(a)}_X\,\frac{\overline{E}_X}{\overline{m} \,c^2}\,.
\end{equation}
Therefore tests of UCR and UFF are not independent, and we can compare
their different qualitative meaning. Since for typical energies
involved we shall have $\overline{E}_X\ll\overline{m} \,c^2$ this
means that $\beta_X\ll\alpha_X$.\footnote{We denote by $\beta_X$ and
  $\alpha_X$ some typical values of the parameters $\beta^{(a)}_X$ and
  $\alpha^{(a)}_X$ for different bodies $(a)$.}  For a given set of
UFF and UCR tests their relative merit is given by
Eq.~\eqref{alphabeta} and is dependent on the model used, i.e. the
type of anomalous energy $E_X$ and its dependence on the used
materials or atoms.\footnote{Imposing that UFF and UCR tests be
    quantitatively equivalent, requires that the
    EEP-violating ``internal'' energy $E_X$ be the total mass, so that
    $\overline{E}_X=\overline{m} \,c^2$. But then $\alpha=\beta$ is a
    universal parameter which can be re-absorbed into the definition
    of $g$. To obtain an effect in this case one needs to invoke the
    multiple Lagrangian formalism of Sec.~\ref{IIIA}.}

For example, let us assume a model in which all types of
electromagnetic energy are coupled in a non-universal way,
i.e. $\beta_\text{EM} \neq 0$ (with all other forms of energies
behaving normally), and where the clock transition is purely
electromagnetic. The UFF test between two materials $(a)$ and $(b)$,
both containing electromagnetic energy, is carried out with an
uncertainty of $\vert\beta^{(a)}_\text{EM} -
\beta^{(b)}_\text{EM}\vert\simeq\vert\beta_\text{EM}\vert \lesssim
10^{-13}$ \cite{Williams,Schlamminger}. On the other hand the UCR test
for a clock of type $(a)$ based on an electromagnetic transition, is
carried out with an uncertainty of $\vert\alpha^{(a)}_\text{EM}\vert
\simeq\vert\alpha_\text{EM}\vert \lesssim 10^{-4}$ \cite{Vessot}. For
macroscopic test bodies the nuclear electromagnetic binding energy
contributes typically to about
$\overline{E}_\text{EM}/(\overline{m}c^2) \simeq 10^{-3}$ of the total
mass \cite{Nordtvedt}, so we may have $\vert\beta_\text{EM}\vert
\simeq 10^{-3}\,\vert\alpha_\text{EM}\vert$, which yields for the UFF
test a limit of about $\vert\alpha_\text{EM}\vert\lesssim 10^{-10}$.
This is still a much more stringent limit than the UCR test
$\vert\alpha_\text{EM}\vert \lesssim 10^{-4}$. So in such a model UFF
tests are significantly more sensitive than UCR tests.

However, that result depends on the particular model used. If we
assume another model in which the nuclear spin plays a role
leading to a non-universal coupling of atomic hyperfine energies,
i.e. $\beta_\text{HF} \neq 0$ (with other forms of energies and
properties of the atom behaving normally), the result is
different. The
  precise mechanism is unknown, but one could imagine an abnormal
  coupling of nuclear spin to gravity, but with the standard coupling
  of the electromagnetic nuclear binding energy, which is spin
  independent, still satisfied. Atomic hyperfine energies are of
order $10^{-24}\,\text{J}$ (corresponding to GHz transition
frequencies) which for typical atomic masses leads to
$\bar{E}_\text{HF}/(\bar{m}c^2)\simeq 10^{-16}$. As a consequence UFF
tests set a limit of only $\vert\alpha_\text{HF}\vert\lesssim 10^{3}$
while UCR tests using hyperfine transitions (e.g. H-masers
\cite{Vessot}) set a limit of about $\vert\alpha_\text{HF}\vert
\lesssim 10^{-4}$. The conclusion is therefore radically different in
this model where UCR tests perform orders of magnitude better than UFF
tests.

To summarize, the two types of tests, UFF and UCR, are
complementary in the modified Lagrangian framework, and need to be
pursued with equal vigor, because depending on the model used
either one of the tests can prove significantly more sensitive
than the other. A corollary is that usual tests of UCR do not need
a violation of Schiff's conjecture to be meaningfully interpreted.
They can perfectly be contrasted with UFF tests within a formalism
satisfying the Schiff conjecture.

Let us now consider the application to atom interferometry. In the
experiment of Refs.~\cite{Peters,Mueller} the ``atom-clock'' that
accumulates a phase is of identical composition to the falling
object (the same atom), hence one has to consistently use the same
value of $\beta^{(a)}_X$ when calculating the trajectories and the
phase difference using the Lagrangian \eqref{Lmodified} inserted
into \eqref{DS}. It is then easy to show that $\Delta\varphi_S=0$
with the above Lagrangian \cite{Storey,Borde01,Wolf}. The
vanishing of $\Delta\varphi_S$ in this case is a general property
of all quadratic Lagrangians and comes from consistently using the
same Lagrangian for the calculation of the trajectories and the
phase shift, as shown in Sec.~\ref{IIB} above. It is related to
the cancellation between the special relativistic effect [last
term in \eqref{Lmodified}] and the gravitational potential energy
[third term in \eqref{Lmodified}], discussed in more detail below.

Then the total phase shift of the atom interferometer is again given
by the light interactions only, which are obtained from the phases at
the interaction points evaluated using the trajectory given by
\eqref{Lmodified}. From Eq.~\eqref{Deltaphiell} one then obtains
\begin{equation}\label{dphibetaX}
\Delta\varphi = \Delta\varphi_\ell =
\bigl(1+\beta^{(a)}_X\bigr)k\,g\,T^2\,.
\end{equation}
We first note that in this class of theories the Compton frequency
plays no role, as $\Delta\varphi_S = 0$. Indeed the mass term
  $-m_0 c^2$ in the Lagrangian \eqref{Lmodified} cancels out and could
  be ignored. Second, we note that
although $\beta^{(a)}_X$ appears in the final phase shift, this is
entirely related to the light phase shift coming from the trajectory
of the atoms, and thus is a measurement of the effective free fall
acceleration $(1+\beta^{(a)}_X)g$ of the atoms, which is given by
Eq.~\eqref{acc}. In Refs.~\cite{Peters,Mueller} the resulting phase
shift is compared to $k\,\tilde{g}\,T^2$ where $\tilde{g}$ is the
measured free fall acceleration of a falling macroscopic corner cube,
i.e. $\tilde{g}=\bigl[1+\beta^{(\text{corner cube})}_X\bigr]g$, also
deduced from Eq.~\eqref{acc}. In this class of theories the experiment
is thus a test of the universality of free fall, as it measures the
differential gravitational acceleration of two test masses (C{\ae}sium
atom and corner cube) of different internal composition, with
precision
\begin{equation}\label{testUFF}
\left|\beta^{(\text{Cs})}_X - \beta^{(\text{corner cube})}_X
\right|\lesssim 7 \times 10^{-9}\,,
\end{equation}
for any kind of internal energy $X$. Note that this expression is
equivalent to the one obtained by MPC in their recent paper
\cite{Hohensee2} (p.~4), but different from the one obtained in
the same paper (p.~3) for UCR tests (see also a footnote in
Sec.~\ref{IB}).

As already mentioned the vanishing of $\Delta\varphi_S$ and thus
the non-sensitivity to the gravitational redshift in the
``atom-clock'' phase is related to the cancellation between the
special relativistic term and the gravitational term when
integrating along the trajectories. More generally, the underlying
fundamental point here is that for a measurement of the
gravitational redshift one wants to determine the non-zero effect
of the gravitational redshift term [third term in \eqref{LGR} or
\eqref{Lmodified}] on the clock phase (atom or classical)
\textit{independently} of the effect of the special relativistic
term [fourth term in \eqref{LGR} or \eqref{Lmodified}] and of
possible other phase shifts (e.g. the light phase $\Delta
\varphi_\ell$). This is generally done by using independent
measurements that provide the knowledge of the trajectories of the
clocks $z(t)$ and their velocities $\dot{z}(t)$. One uses that
knowledge to subtract the time dilation and light phase shifts
from the overall measured shift $\Delta \varphi$. The remaining
gravitational redshift term can then be compared to the GR
prediction.

As an example, in the Pound-Rebka experiment \cite{Pound,PoundSnider}
the two clocks were fixed to the Earth's surface at the top and bottom
of the tower, which imposes $z(t)=\text{const}$ and
  $\dot{z}(t)=0$. As a result the first-order Doppler shift was zero
(implying $\Delta \varphi_\ell = 0$) and the special relativistic term
in the action phase shift also.\footnote{Actually, in the real
    experiment the measured signal is given by a first-order Doppler
    shift which is applied to compensate for the gravitational
    redshift.} The measured frequency or phase difference was then
entirely due to the gravitational term and could be compared to the GR
prediction, clearly favoring GR. In modern rocket or satellite
experiments \cite{Vessot,Cacciapuoti} the trajectories of the clocks
are measured independently using radio and/or laser ranging which
allows calculation and correction of the special relativistic and
light phase-shifts, and thus an unambiguous measurement of the
gravitational redshift. In contrast, in the atom-interferometry
experiment the trajectories are not measured independently but
theoretically derived from the Lagrangian and initial conditions, thus
making a ``pure'' measurement of the gravitational term
impossible. One might even argue that in such an experiment the
independent determination of the wave packet trajectories (e.g. by
high resolution imagery) would be impossible as it would destroy the
interference at the interferometer exit according to one of the
fundamental tenets of quantum mechanics.

\section{Conclusion}\label{IV}

We have shown that it was possible to provide answers to the question
asked in the title of this article provided that the alternative
theory or theoretical framework is carefully specified.  To this aim,
after having proved in Sec.~\ref{II} that within GR atom
interferometers are insensitive to the gravitational redshift, we have
considered two classes of alternative theories to GR.

The first class of models is based on a modified Lagrangian (see
Sec.~\ref{IIIB}). In this class one finds most alternative
approaches that postulate a non-universal coupling between gravity
and other fields of the standard model of particle physics or its
extensions (see e.g.
\cite{Damour,LightLee,Blanchet,Koste1,Koste2}), including some
versions of string theory, brane scenarios, parameterized
frameworks, etc. This class of models is consistent with Schiff's
conjecture and represents the most natural framework for analyzing
and comparing the various consequences of the violation of the
equivalence principle. In this class of models, tests of the
universality of free fall (UFF) and of the universality of clock
rates (UCR) are related, with the quantitative relationship
depending on the particular theory used, and the corresponding
nuclear and atomic models. We have shown that in this class of
models the free propagation phase shift $\Delta \varphi_S=0$ and
therefore the Compton frequency plays no role; the interferometer
cannot be viewed as measuring the cumulated phase-difference of
the ``Compton-clocks''. Instead the final phase difference is
given by the light interactions and thus measures the free fall
acceleration of the atoms compared to the free fall acceleration
of a corner cube of different composition, akin to classical UFF
tests. The atom gravimeter experiments are thus tests of the weak
equivalence principle or universality of free fall at the
$7\times10^{-9}$ accuracy level, see Eq.~\eqref{testUFF}. As such
they are less accurate than other UFF tests
\cite{Williams,Schlamminger} but still interesting as they are the
best tests of this kind which compare the free fall of quantum
objects to that of classical test masses.

The second class of models relies on multiple Lagrangians
(Sec.~\ref{IIIA}). Within this class, the atom interferometry
experiment can be seen as a test of the gravitational redshift, in
line with the claims of MPC \cite{Mueller}, but at the expense of
extreme conceptual difficulties: the basic principles of quantum
mechanics, such as the Feynman path integral formulation and
Schr\"odinger's equation, are affected, as well as energy conservation
and the Schiff conjecture. This is because two Lagrangians are used in
this class of models, one for the trajectory of the atoms and another
one for the phase of the associated matter field. In addition, these
models pose a fundamental problem for calculating phases in atom
interferometry tests, since the standard formula for phase shifts is
based on principles of quantum mechanics which are violated in this
class of alternative theories.

We conclude that the general statement of MPC according to which
the atom interferometer measures or tests the gravitational
redshift at the Compton frequency is incorrect. Instead, the
interpretation of the experiment needs to be considered in the
light of alternative theories or frameworks. Although one can
consider particular alternative frameworks in which the statement
of MPC could make sense, such frameworks raise unacceptable
conceptual problems which are not at the moment treated in a
satisfactory manner. In particular, they break the fundamental
principles of quantum mechanics which are used for calculating
matter wave phases. In most common and plausible theoretical
frameworks the atom interferometry experiment tests the
universality of free fall with the Compton frequency being
irrelevant.

\vspace{0.3cm} \textit{Note added}: After completion of the
  present work, several independent analysis have appeared
  \cite{Samuel,Giulini} supporting our views and consistent with our
  conclusions.

\appendix

\section{Phase shift in neutron interferometry}\label{appA}

In this Appendix we show that the phase shift induced by gravity in
neutron interferometers is given by an expression similar to that in
atom interferometry. In consistency with the analogy between neutron
and atom interferometry pointed out in Ref.~\cite{BordeBook}, the
gravitational phase shift is independent of the Compton frequency also
for the neutrons.

\begin{figure}
\begin{center}
\includegraphics[width=12cm,angle=0]{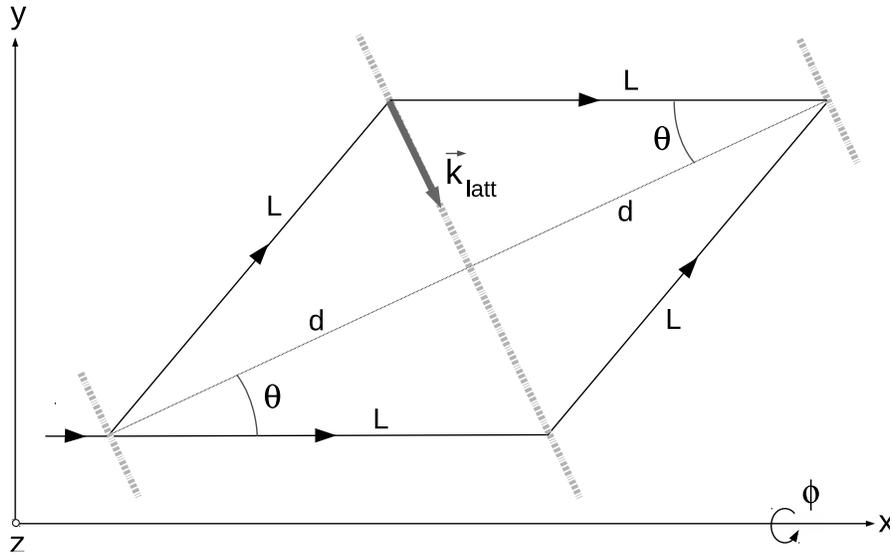}
\caption{Geometry of the neutron interferometer, simplified from
  Fig.~1 of Ref.~\cite{COW}. Thick dotted lines represent the
  scattering crystal planes, thin solid lines the neutron
  trajectories.}
\label{fig2}
\end{center}
\end{figure}

Neutron interferometers have been first used for the measurement of
the gravitational acceleration in 1975 \cite{COW}. A neutron beam is
Bragg scattered by three silicon crystal planes and forms an
interferometer as shown in Fig.~\ref{fig2}.
The phase shift at the exit of the interferometer is given in
Ref.~\cite{COW} to leading order as
\begin{equation}\label{COWshift}
  \Delta\varphi=\frac{\lambda_\text{dB}M^2\,d^2\,g}{\pi\hbar^2}
  \tan\theta \,\sin\phi\,,
\end{equation}
where $\lambda_\text{dB}$ and $M$ are the neutron de Broglie
wavelength and mass respectively. Using the definition of the de
Broglie wavelength $2\pi\hbar/\lambda_\text{dB}=Mv$, where $v$ is the
velocity of the neutrons, Eq.~\eqref{COWshift} can be rearranged to
\begin{equation}\label{COWshift2}
  \Delta\varphi=\frac{4\pi g}{\lambda_\text{dB}}T^2
  \sin\theta\cos\theta\,\sin\phi\,,
\end{equation}
with $T\equiv L/v$. The well known Bragg condition for the scattering
at the crystal planes is
\begin{equation}\label{Bragg}
\lambda_\text{dB}=2a\,\sin\theta \,,
\end{equation}
with $a$ the atomic spacing in the crystal lattice. Substituting
Eq.~\eqref{Bragg} into Eq.~\eqref{COWshift2} one easily obtains
\begin{equation}\label{COWshift3}
  \Delta\varphi=k_\text{latt}\,g\,T^2\,\cos\theta \sin\phi
  = \bm{k}_\text{latt}\cdot\bm{g} \,T^2\,,
\end{equation}
where $k_\text{latt}\,\cos\theta \sin\phi = (2\pi/a)\,\cos\theta
\sin\phi$ is the projection of the lattice wave vector
$\bm{k}_\text{latt}$ onto the direction of the gravitational field
$\bm{g}$.

We note that Eq.~\eqref{COWshift3} is formally identical to
Eq.~\eqref{Deltaphitotal} with the lattice wave vector playing the
role of the laser wave vector. In both cases, the expression for the
phase shift is independent of the mass of the atoms (neutrons) or the
associated Compton frequency. In both cases, the interferometer is to
be understood as measuring the free fall of the atoms (neutrons) using
the laser (crystal lattice) as a ``ruler''.

\section{Effect of high-order gravity gradients}\label{appB}

As we have seen, for any quadratic Lagrangian the difference of
classical actions \eqref{DS} is zero or reduces to the
contribution of internal energies. In this Appendix, although this
has already
  been presented in much greater detail in Ref.~\cite{Dimopoulos}, we
estimate the magnitude of the effect of including cubic terms in the
Lagrangian of GR, due to second-order gravity gradients, i.e.
\begin{equation}\label{GRcubic}
L_\text{GR}(z,\dot{z})=-mc^2+\frac{GMm}{r_\oplus}+mg\Bigl[-z +
\frac{z^2}{r_\oplus} -
\frac{z^3}{r_\oplus^2}\Bigr]+\frac{1}{2}m\dot{z}^2\,.
\end{equation}
For such cubic Lagrangian we expect that the difference of classical
actions will no longer be zero. On dimensional grounds we expect that
the result should then be given by\footnote{To \eqref{resC} we should
  also add the contribution of $\Delta\varphi_\ell$.}
\begin{equation}\label{resC}
\Delta \varphi_S = \frac{\Delta S_\text{cl}}{\hbar} =
C \,k \,g \,T^2 \left(\frac{\hbar \,k \,T}{m
\,r_\oplus}\right)^2
+ \Delta\varphi_{gg'}\,,
\end{equation}
where $C$ denotes some global coefficient of the order of one. By
integrating the Lagrangian \eqref{GRcubic} along the classical paths
one obtains $C=1/4$. We note that for the Lagrangian \eqref{GRcubic}
the interferometer closes up only when we adjust the time $T'$ to a
value that is different from $T$. So the interferometer is no longer
symmetric and there is a contribution coming from the difference of
internal energies $gg'$ of the atoms, given by
\begin{equation}
\Delta \varphi_{gg'} = \omega_{gg'} \frac{g T^3}{r_\oplus}\,.
\end{equation}

The first term in \eqref{resC} gives a non zero contribution which
could be interpreted as a redshift effect by analogy with classical
clock experiments. However, we insist that in order to compute
correctly the phase shift one would have to revisit the derivation
made in Sec.~\ref{IIA} for the case of cubic and higher-order
Lagrangians. If we assume that at least the order of magnitude given
by the first term in \eqref{resC} is correct, the effect is extremely
small. With $\hbar k/m\simeq3\,\mathrm{mm}/\mathrm{s}$,
$T\simeq0.1\,\mathrm{s}$ and $r_\oplus=6400\,\mathrm{km}$, it gives a
relative correction to the main effect (i.e. $\Delta\varphi_\ell=k g
T^2$) of the order of $2\times 10^{-21}$, much too small to be
measured in the atom interferometry experiments
\cite{Peters,Mueller,refsyrte} whose current precision is about
$7\times 10^{-9}$. Therefore we conclude that the redshift effect in
atom interferometry, if it exists at all, is very small and currently
undetectable.

\end{document}